\documentclass[12pt]{article}

\newtheorem{prop}{Proposition}
\newtheorem{theorem}[prop]{Theorem}
\newtheorem{lemma}[prop]{Lemma}

\usepackage{epsfig}

\begin{document}

\title{Finite gap integration of a discrete Euler top}
\author{B.\,Lorbeer\footnote{E--mail: lorbeer@math.tu-berlin.de}}
\date{Fachbereich Mathematik,
Technische Universit\"at Berlin, Str. 17 Juni 135, 10623 Berlin,
Germany}
\maketitle

\begin{abstract}
In [1] new discretizations of the Euler top have been found. They
can be discribed with a Lax pair with a spectral parameter on
an elliptic curve. This is used in this paper to perform a finite
gap integration.
\end{abstract}

\newpage 

\section{Introduction}

The motion of a rigid body which is not submitted to gravitational force,
this is known as the Euler case, is one of the few tops which are completely
integrable. These systems have been studied by lots of famous mathematicians
like Jacobi and Klein to mention only two.

A new approach to integrable systems in general (finite and infinite
dimensional) is provided by the method of finite gap integration which
uses Lax pairs with a spectral parameter.
This has been introduced by the work of Novikov,
Dubrovin, Matveev, Its, Kriechever, and others. Once such a Lax
pair is found, one can start to use algebro-geometric methods to describe
the system. Accounts of what has been acheived up to now can be found
in [2]. 

In the last ten to twenty years completely integrable discrete
systems gained more and more attention. Among the early works
in this field I only mention [5], [6], [7], [8]. A collection
of new articles relevant within this context is [9].
[7] is especially important for the present paper, since it is
the first one to give a description of the discrete Euler top
with the help of a Lax pair. Here the spectral parameter varies
on ${\bf C}P^1$. The ideas put down there where elaborated further
in [10] and [11].

In [1] we studied another Lax pair for the discrete Euler top 
where the spectral parameter varies on an elliptic curve. This makes
the finite gap integration more difficult. To my knowledge
the first solution of such a system was given by A.I. Bobenko in
[12]. The present paper can be seen as an addendum to [1]. It describes
the finite gap integration for the system introduced there.

\section{A Discrete Euler Top}

In [1] a discrete version of the Euler top was studied. These
notes now are concerned with the finite gap integration of this system.
Let's recall the notation:

The top itself is given by the inertia tensor (in diagonal form):

\begin{equation}\label{I}
{\bf I}=diag(A,B,C)\;, A>B>C>0.
\end{equation}

\noindent The vectors ${\bf M},{\bf \Omega}$ denote the angular momentum and the
angular velocity respectively. That is ${\bf M}={\bf I \Omega}$ and the
continuous Euler top is given by the well known Euler equation in ${\bf R}^3$:

\begin{equation}\label{evol}
{\bf \dot{M}}={\bf M}\times{\bf \Omega}.
\end{equation}

\noindent The discrete version of the Euler top mentioned above is given by a Lax pair depending on
a spectral parameter varying on a torus.
Let:

\begin{eqnarray}\label{ws}
w_1(u) &=& \rho\,\frac{1}{{\rm sn}(u,\kappa)}\;,\nonumber\\
w_2(u) &=& \rho\,\frac{{\rm dn}(u,\kappa)}{{\rm sn}(u,\kappa)}\;,\\
w_3(u) &=& \rho\,\frac{{\rm cn}(u,\kappa)}{{\rm sn}(u,\kappa)}\;.\nonumber
\end{eqnarray}

\noindent Here sn, cn, dn are the Jacobi elliptic functions and the parameter
$\rho$ and the module $\kappa$ are defined by

\begin{eqnarray}\label{rokappa}
\rho^2&=&\frac{C-A}{AC}\;,\nonumber\\
\kappa^2&=&\frac{C(A-B)}{B(A-C)}\;.
\end{eqnarray}

\noindent Next let ${\bf V}=(V_1,V_2,V_3)^t$ be a vector in
${\bf R}^3$, $\sigma_k(k=1,2,3)$ the Pauli matrices
and finally let $u_0$ be such that

\begin{equation}\label{u0}
w_1(u_0)=A^{-\frac{1}{2}}\; , \quad w_2(u_0)=B^{-\frac{1}{2}}\; , \quad w_3(u_0)=C^{-\frac{1}{2}}\; ,
\end{equation}

\noindent (such a point exists because of the special choice of the parameters $\rho,\kappa$ ).
Now we can define the Lax-pair:

\begin{eqnarray}\label{MV}
M&=&\frac{1}{2i}\sum_{k=1}^3 M_kw_k(u)\sigma_k\; ,\nonumber\\
\mathcal{V}&=&{\bf 1}+\frac{h}{2i}\sum_{k=1}^3V_kw_k(u_0-u)\sigma_k\;.
\end{eqnarray}

\noindent The discrete evolution is given by the Lax-equation

\begin{equation}\label{lax}
M_{n+1}(u)={\mathcal V}_{n}(u)M_n(u){\mathcal V}^{-1}_n(u)\; ,
\end{equation}

\noindent which will sometimes also be written as

\begin{equation}\label{laxHat}
\hat M(u)={\mathcal V}(u)M(u){\mathcal V}^{-1}(u)\;.
\end{equation}

\noindent These are the compatibility conditions for the following system:

\begin{equation}\label{auxil}
\left|
\begin{array}{rcl}
M\Phi&=&\Phi\breve \mu\\
\hat \Phi&=&{\mathcal V}\Phi
\end{array}
\right|
\end{equation}

\noindent Here $\breve \mu=diag(\mu_1,\mu_2)$ where $\mu_1$ and $\mu_2$ are
the eigenvalues. In [1] it has been shown that (\ref{lax}) is equivalent
to a discrete Euler top given by

\begin{equation}\label{top}
\hat M-M=\frac{\gamma h}{4}(M+\hat M)\times {\bf I}^{-1}(M+\hat M).
\end{equation}

\noindent Here $h$ is the discrete parameter and $\gamma$ a real number which
tends to $1$ when $h$ tends to zero. See [1] for more
considerations on $\gamma$ (e.g. how $\gamma$ has to look like so
that the discrete map becomes a Poisson map). This
paper is concerned with finding explicit solutions for this system
with the help of Prym theta functions and abelian differentials. The
methods used here are similar to those in [2].

\section{The Spectral Curve}

To perform the finite gap integration, we first have to get the
spectral curve. From (\ref{lax}) it is clear that for each fixed $u$
the eigenvalues of $M(u)$ are independent of $n$. So the solutions of

\begin{equation}\label{curve}
{\rm det}(M(u)-\mu{\bf 1})=0
\end{equation}

\noindent are constant. Using the identities

\begin{eqnarray}\label{wsquare}
w_2^2-B^{-1}&=&w_1^2-A^{-1}\;,\nonumber\\
w_3^2-C^{-1}&=&w_1^2-A^{-1}\; ,
\end{eqnarray}

\noindent (\ref{curve}) is equivalent to

\begin{equation}\label{curve2}
\mu^2=\frac{-1}{4}(M^2w^2_1+E-A^{-1}M^2),
\end{equation}

\noindent where $E=\frac{M_1^2}{A}+\frac{M^2_2}{B}+\frac{M^2_3}{C}$.

Let's denote the torus on which the Jacobi elliptic functions are
defined by ${\bf T}$ (the two generators for the lattice of the torus
are given by $4K$ and $4iK'$ where $K$ is the complete elliptic
integral of the modulus $\kappa$ and $K'$ is the complete elliptic
integral of the complementary modulus; both are real in this case).
Furthermore let

\begin{eqnarray}
\lambda :{\bf T}\to {\bf T},\; u\to u+2K&&\label{lambda}\\
\epsilon :{\bf T}\to {\bf T},\; u\to u+2iK'.&&\label{tau}
\end{eqnarray}

\noindent If $u$ is a point on the torus the set $\{u,\lambda u,\epsilon u,\lambda
\epsilon u\}$ will be denoted by $\langle u\rangle$. Next, $w_1^2$ is invariant
w.r.t. $\lambda$ and $\epsilon$, so the set of solutions of (\ref{curve2})
is invariant w.r.t. $(u,\mu)\to (u+2K,\mu)$ and $(u,\mu)\to
(u+2iK',\mu)$ which henceforth will also be denoted by $\lambda$ and
$\epsilon$ respectively.

\begin{lemma}
The set of solutions of(\ref{curve2}) can be made into a Riemann
surface.
\end{lemma}

{\bf Proof}: Everywhere except at the points $\langle u_b\rangle$ with $\mu^2(u_b)=0$
and at the points $\langle 0\rangle$ , $u$ is a chart. At the branch points $\langle u_b\rangle$

\begin{equation}
\mu^2=[-\frac{M^2}{2}w_1(u_b)w'_1(u_b)](u-u_b)+\dots, u\to u_b
\end{equation}

\noindent Now let's presume for the moment that $w_1(u_b)w'_1(u_b)\neq 0$. Then
$\mu$ is obviously a chart.

Next, delete the line $\{(0,z)|z\in{\bf C}\}$ from ${\bf T}\times{\bf C}$,
obtaining, say, $({\bf T}\times {\bf C})^o$. Let $B$ be a small disk
in ${\bf C}$ centered at zero, $B^o$ the punctured disk, and glue
$B\times {\bf C}$ to $({\bf T}\times {\bf C})^o$ via the following map
$m$:

\begin{equation}\label{m}
m:B^o\times{\bf C}\to ({\bf T}\times {\bf C})^o,\; (t,\nu)\to(u,\mu)=(t,
\frac{\nu}{t})
\end{equation}

\noindent In the new coordinates the curve (\ref{curve2}) is given by

\begin{eqnarray}
\nu^2&=&\frac{-1}{4}t^2(M^2w_1^2(t)+E-A^{-1}M^2)\nonumber\\
     &=&\frac{-1}{4}M^2\rho^2+O(t),t\to 0.
\end{eqnarray}

\noindent Thus, adding two points at $(t=0,\nu=\pm \frac{M^2\rho^2}{4})$, the set
of solutions of (\ref{curve2}) is compactified to a compact
Riemann surface with $u$ a chart at the points $\langle 0\rangle$.\hspace*{\fill}{\bf qed.}
\vspace{.5 cm}

This compact curve is called the spectral curve and will be referred to
by the symbol $C$. It is invariant w.r.t. $\lambda$ and $\epsilon$. The
exchange of sheets will be denoted by $\pi$:

\begin{equation}\label{pi}
\pi :C\to C, \; (u,\mu)\to (u,-\mu).
\end{equation}

\noindent For each of the maps $\lambda, \epsilon$, and $\pi$ we have natural maps
to the belonging quotient spaces, which are all compact Riemann
surfaces, too, and these maps will be denoted by $pr_\lambda,
pr_\mu, pr_\pi$, respectively. At times, we will also make use of the
following map:

\begin{equation}\label{chi}
\chi : C\to C,\; (u,\mu)\to (-u,-\mu).
\end{equation}

\noindent Next, the locations of the branch
points have to be described better. They are given by:

\begin{equation}
-w_1^2+A^{-1}-\frac{E}{M^2}=0
\end{equation}

\begin{figure}
\begin{center}

\begin{picture}(0,0)%
\epsfig{file=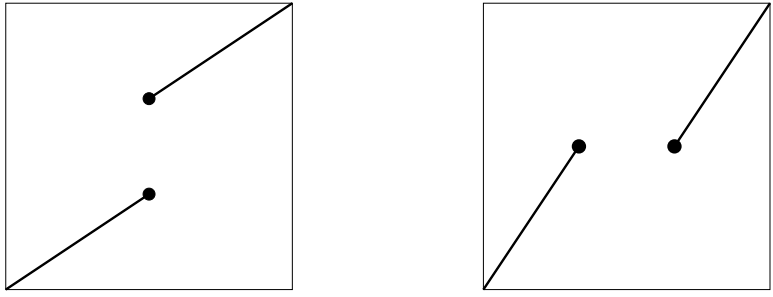}%
\end{picture}%
\setlength{\unitlength}{1973sp}%
\begingroup\makeatletter\ifx\SetFigFont\undefined%
\gdef\SetFigFont#1#2#3#4#5{%
  \reset@font\fontsize{#1}{#2pt}%
  \fontfamily{#3}\fontseries{#4}\fontshape{#5}%
  \selectfont}%
\fi\endgroup%
\begin{picture}(9666,3666)(1168,-5194)
\end{picture}

\caption{from left to right: $\frac{E}{M^2}>B^{-1}$ and $\frac{E}{M^2}<B^{-1}$} \label{EDiffers}
\end{center}
\end{figure}

\noindent Using the definition of $\rho$, $\kappa$, and of sn as the inverse
of the first elliptic integral, we arrive at the following 
situation: There are eight branch points, two of them ($u_{b1}$ and
$u_{b2}$) in the small torus ${\bf T}/\lambda/\epsilon$
(the others are obtained applying
the maps $\lambda , \epsilon$, and $\lambda\epsilon$). For
$\frac{E}{M^2}>\frac{1}{B}$ they are situated on the line
$K\to K+2iK'$, where $\bar u_{b1}=u_{b2}$. For $\frac{E}{M^2}<\frac{1}{B}$
they are situated on the line $iK'\to 2K+iK'$ and are symmetric
over the point $K+iK'$. When $E=\frac{M^2}{C}$ (this corresponds to
the stable fix points $(0,0,\pm |M|)$), the eight branch
points degenerate to four points given by $\langle K\rangle$. For $E=\frac{M^2}{A}$
(corresponding to the stable fix points $(\pm |M|,0,0)$) we get $\langle iK'\rangle$
and finally for $E=\frac{M^2}{B}$ (corresponding to the separatrix
of the unstable fix points $(0,\pm |M|,0)$) we arrive at
$\langle K+iK'\rangle$.
Next we find the appropriate cuts.
Using the new coordinates $(t,\nu)$ at $u=0$ shows that on
a path on $C$ which projects to a straight line through $u=0$ the function
$\mu$ changes sign, i.e. starting at $(u,\mu)$ we end up at $(u,-\mu)$ ($C$
is invariant w.r.t. the map $(u,\mu)\to (-u,\mu)$).
Thus, going from, say, $iK'$ to $3iK'$ and from $K$ to $3K$, an
odd number of cuts should be crossed. Hence an appropriate picture
of $C/\lambda/\epsilon$ would be figure (\ref{EDiffers}).

\section{The Divisors Of The Baker-Akhiezer Functions}
The eigenvectors of the matrices $M_n$ (see (\ref{auxil})) can be
understood as functions
on the spectral curve: just attach to the point $(u,\mu)$
the eigenvector $\phi$ of $M_n(u)$ with eigenvalue $\mu$
(they are, of course, determined only up to
multiplication by an arbitrary function on $C$, so an appropriate
normalization will be introduced (see e.g. [3])).

The idea of finite-gap integration is to find as many information as
possible about the divisors of the eigenvector functions $\phi_n$
and then to reconstruct these
functions and from them finally the matrices $M_{n}$ and the vectors
${\bf M}_n$. The reason why this is possible is that the considered
functions are meromorphic on a $compact$ Riemann surface, hence
strong tools like the Riemann-Roch theorem can be used.

First we consider the eigenvector function $\phi_0(u,\mu)$ of the
matrix $M_{n=0}$ on the spectral curve. It is defined at each point
up to a multiplicative number and so it is possible to take the
normalized vector function:

\begin{equation}\label{phi0}
\phi_0(u,\mu)=\left(\begin{array}{c}
                    1\\
                    \phi_{0_2}
                    \end{array}\right).
\end{equation}

\noindent To find the zeros and poles of $\phi_{0_2}$ we introduce the following
abbreviations:

\begin{eqnarray}
M_0(u)&=&\left(\begin{array}{cc}
               \alpha & \beta\\
               \delta & -\alpha
               \end{array}\right),\label{abbr}\\
\mu^2&=&\alpha^2+\beta\delta.\label{muSquare}
\end{eqnarray}

\begin{lemma}
Let $\breve u\in {\bf T}$ be such that $\beta(\breve u)=0$. Then define

\begin{equation}\label{P}
P(\breve u):=(\breve u,\alpha(\breve u))+ \lambda
(\breve u,\alpha(\breve u))
+\chi(\breve u,\alpha(\breve u))+\lambda \chi(\breve u,\alpha(\breve u)).
\end{equation}

\noindent With this notation the pole divisor of $\phi_{0_2}$ is given
by $P(\breve u)$ and the zero divisor by $\epsilon P(\breve u)$, i.e.:

\begin{equation}\label{phi0Divisor}
(\phi_{0_2})=\epsilon P(\breve u)-P(\breve u).
\end{equation}

\end{lemma}

{\bf Proof}: We have 

\begin{equation}\label{phi02}
\phi_{0_2}=\frac{\mu-\alpha}{\beta},
\end{equation}

\noindent which is a meromorphic function on $C$.
It can only have poles where $\beta=0$ or
$\mu-\alpha=\infty$.
Over $\breve u$ we have $\mu
=\pm \alpha$, so there is exactly one pole over
$\breve u$ at $(\breve u,+\alpha)$ ($\mu =\alpha =0$ only
in the uninteresting case $E=\frac{M^2}{C}$, since $(\alpha)_{zero}
=(\omega_3)_{zero}=K+\lambda K+\epsilon K+\lambda \epsilon K.$)
Now the functions $w_k \,(k=1,2,3)$ have first order poles at the points
$\langle 0\rangle$. Hence $\beta$ has four poles and thus also four zeros
in ${\bf T}$.
This gives four poles for $\phi_{0_2}$ in $C$. Recalling the periodicity of
the $w_k$:

\begin{equation}\label{wPeriods}
\begin{array}{rcrrcr}
w_1(\lambda u)&=&-w_1(u), & w_1(\epsilon u)&=&w_1(u),\\
w_2(\lambda u)&=&-w_2(u), & w_2(\epsilon u)&=&-w_2(u),\\
w_3(\lambda u)&=&w_3(u), & w_3(\epsilon u)&=&-w_3(u),
\end{array}
\end{equation}

\noindent and the fact that all $w_k$ are odd, i.e $\phi_{0_2}(p)=
\phi_{0_2}(\chi p)$, we see that these four poles are given
by $P(\breve u)$.
The function $\mu-\alpha$ can only have poles over the points
$\langle 0 \rangle$ where all $w_k$ have a pole, i.e. $\beta$ too, so the
four poles above are the only ones.

Accordingly, $\phi_{0_2}$ has four zeros given by $\delta=0$ (see(\ref{muSquare})).
From (\ref{wPeriods}) it follows that $\beta(u+2iK')=
\delta(u)$ and $\alpha(u+2iK')=-\alpha(u)$, thus the zero
divisor is given by $\epsilon P(\breve u)$. \hspace*{\fill}{\bf qed.}
\vspace{.5 cm}

The next step is the discription of the discrete evolution. We will,
for every $n\in {\bf N}$, find enough properties of $\phi_n=
{\mathcal V}_{n-1}\dots{\mathcal V}_0\phi_0$, which define this
n-th eigenvector function uniquely.

\begin{figure}
\begin{center}

\begin{picture}(0,0)%
\epsfig{file=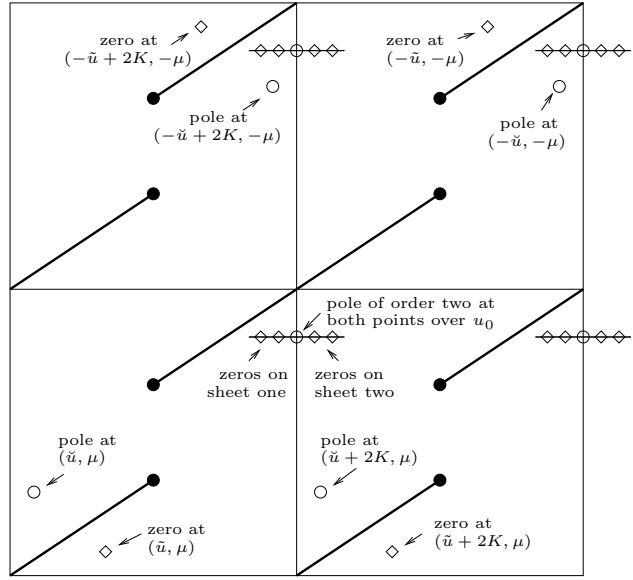}%
\end{picture}%
\setlength{\unitlength}{1973sp}%
\begingroup\makeatletter\ifx\SetFigFont\undefined%
\gdef\SetFigFont#1#2#3#4#5{%
  \reset@font\fontsize{#1}{#2pt}%
  \fontfamily{#3}\fontseries{#4}\fontshape{#5}%
  \selectfont}%
\fi\endgroup%
\begin{picture}(7845,7266)(2368,-7594)
\put(5026,-5086){\makebox(0,0)[lb]{\smash{\SetFigFont{6}{7.2}{\rmdefault}{\mddefault}{\updefault}zeros on}}}
\put(6226,-5086){\makebox(0,0)[lb]{\smash{\SetFigFont{6}{7.2}{\rmdefault}{\mddefault}{\updefault}zeros on}}}
\put(6226,-5311){\makebox(0,0)[lb]{\smash{\SetFigFont{6}{7.2}{\rmdefault}{\mddefault}{\updefault}sheet two}}}
\put(6376,-4186){\makebox(0,0)[lb]{\smash{\SetFigFont{6}{7.2}{\rmdefault}{\mddefault}{\updefault}pole of order two at}}}
\put(6376,-4411){\makebox(0,0)[lb]{\smash{\SetFigFont{6}{7.2}{\rmdefault}{\mddefault}{\updefault}both points over $u_0$}}}
\put(4126,-7036){\makebox(0,0)[lb]{\smash{\SetFigFont{6}{7.2}{\rmdefault}{\mddefault}{\updefault}zero at}}}
\put(7726,-6961){\makebox(0,0)[lb]{\smash{\SetFigFont{6}{7.2}{\rmdefault}{\mddefault}{\updefault}zero at}}}
\put(3526,-886){\makebox(0,0)[lb]{\smash{\SetFigFont{6}{7.2}{\rmdefault}{\mddefault}{\updefault}zero at}}}
\put(4126,-7261){\makebox(0,0)[lb]{\smash{\SetFigFont{6}{7.2}{\rmdefault}{\mddefault}{\updefault}$(\tilde u,\mu)$}}}
\put(7726,-7186){\makebox(0,0)[lb]{\smash{\SetFigFont{6}{7.2}{\rmdefault}{\mddefault}{\updefault}$(\tilde u+2K,\mu)$}}}
\put(7126,-886){\makebox(0,0)[lb]{\smash{\SetFigFont{6}{7.2}{\rmdefault}{\mddefault}{\updefault}zero at}}}
\put(7126,-1111){\makebox(0,0)[lb]{\smash{\SetFigFont{6}{7.2}{\rmdefault}{\mddefault}{\updefault}$(-\tilde u,-\mu)$}}}
\put(3076,-1111){\makebox(0,0)[lb]{\smash{\SetFigFont{6}{7.2}{\rmdefault}{\mddefault}{\updefault}$(-\tilde u+2K,-\mu)$}}}
\put(6301,-5911){\makebox(0,0)[lb]{\smash{\SetFigFont{6}{7.2}{\rmdefault}{\mddefault}{\updefault}pole at}}}
\put(6301,-6136){\makebox(0,0)[lb]{\smash{\SetFigFont{6}{7.2}{\rmdefault}{\mddefault}{\updefault}$(\breve u+2K,\mu)$}}}
\put(8551,-1936){\makebox(0,0)[lb]{\smash{\SetFigFont{6}{7.2}{\rmdefault}{\mddefault}{\updefault}pole at}}}
\put(4651,-1861){\makebox(0,0)[lb]{\smash{\SetFigFont{6}{7.2}{\rmdefault}{\mddefault}{\updefault}pole at}}}
\put(3001,-5911){\makebox(0,0)[lb]{\smash{\SetFigFont{6}{7.2}{\rmdefault}{\mddefault}{\updefault}pole at}}}
\put(8401,-2161){\makebox(0,0)[lb]{\smash{\SetFigFont{6}{7.2}{\rmdefault}{\mddefault}{\updefault}$(-\breve u,-\mu)$}}}
\put(3001,-6136){\makebox(0,0)[lb]{\smash{\SetFigFont{6}{7.2}{\rmdefault}{\mddefault}{\updefault}$(\breve u,\mu)$}}}
\put(4876,-5311){\makebox(0,0)[lb]{\smash{\SetFigFont{6}{7.2}{\rmdefault}{\mddefault}{\updefault}sheet one}}}
\put(4201,-2086){\makebox(0,0)[lb]{\smash{\SetFigFont{6}{7.2}{\rmdefault}{\mddefault}{\updefault}$(-\breve u+2K,-\mu)$}}}
\end{picture}

\caption{possible divisor of $\phi_{2_1}$ (to get the divisor of  $\phi_{2_2}$
use $\epsilon$ on the four ``separate'' (away from $\left< u_0\right>$)
zeros} \label{divisorPicture}
\end{center}
\end{figure}

The divisors of $\phi_{n_1}$, $\phi_{n_2}$
are given as follows:

\begin{prop}\label{divisorProp}
Let $U$ be a small neighborhood of $u_0$ in ${\bf T}$ such that $pr_\pi^{-1}(U)$
consists of two nonempty open sets called ``the sheets near $u_0$''.
Further, let $\phi_0$ be the initial eigenvector with $(\phi_{0_2})=
\epsilon P(\breve u)-P(\breve u)$ given by the initial data.
For $h$ small enough, there exist for each $n\in {\bf N}$ a point
$\tilde u_n$ in ${\bf T}$ and two points $p_{n_1}$ and $p_{n_2}$
on $C$, $p_{n_1},p_{n_2}\in pr_\pi^{-1}U$, with the following properties:

\begin{equation}\label{pprop}
\begin{array}{ll}
-&\Im (pr_\pi\, p_{n_1})=\Im(pr_\pi\, p_{n_2})=\Im (u_0)\\
-&pr_\pi\, p_{n_1}=2u_0-pr_\pi \,p_{n_2}\\
-&\mbox{$p_{n_1}$ and $p_{n_2}$ lie in different sheets,}
\end{array}
\end{equation}

from which the divisors of $\phi_{n_1}$, $\phi_{n_2}$ are
constructed as follows:

\begin{eqnarray}
(\phi_{n_1})&=&\sum^n_{l=1}B_l-P(\breve u)+P(\tilde u_n),\label{phiDivisor1}\\
(\phi_{n_2})&=&\sum^n_{l=1}B_l-P(\breve u)+\epsilon P(\tilde u_n),\label{phiDivisor2}\\
B_n&:=&E_n+\lambda E_n+\epsilon E_n +\lambda\epsilon E_n,\label{b}\\
E_n&:=&-(u_0,\mu)-(u_0,-\mu)+p_{n_1}+p_{n_2}.\label{e}
\end{eqnarray}

\noindent If $h$ tends to zero, we get
$\lim_{h\to 0}pr_\pi \,p_{n_k}=u_0\; (k=1,2)$. See figure (\ref{divisorPicture}).

Conversely, for each given set of divisors $\{B_n|n\in{\bf N},\;n>0\}$,
of the form (\ref{b}) with $pr_\pi \,p_{n_1}-u_0$ small, there
is a sequence of vector functions
$\phi_n$ with divisors (\ref{phiDivisor1}), (\ref{phiDivisor2}), which satisfy (\ref{auxil}).
\end{prop}

{\bf Proof}: First we note that multiplication with ${\mathcal V_0}$
adds eight poles to each component of $\phi_0$, one pole at each point
over each element of $\langle u_0\rangle$. We prove this for the
coefficient $\phi_{0_1}$, for the second the procedure is similar.

Recall that

\begin{equation}\label{poleProof1}
\frac{\pm\mu-\alpha}{\beta}(u_0)=\frac{\pm\frac{i}{2}\sqrt{E}-
\frac{1}{2i}\frac{M_3}{\sqrt{C}}}{\frac{1}{2i}(
\frac{M_1}{\sqrt{A}})-i\frac{M_2}{\sqrt{B}})},
\end{equation}

\noindent and

\begin{equation}\label{poleProof2}
V_kw_k(u_0-u)=(\rho V_k)(u_0-u)^{-1}+O(1),\;u\to u_0,\; (k=1,2,3),
\end{equation}

\noindent and for $h\to 0$ we have ${\bf V}\to (\frac{M_1}{\sqrt{A}},
\frac{M_2}{\sqrt{B}}, \frac{M_3}{\sqrt{C}})^t$
(see [1]).

Hence, if no new poles appear, we must have:

\begin{eqnarray*}\label{poleProof3}
\frac{M_3}{\sqrt{C}}+(\frac{M_1}{\sqrt{A}}
-i\frac{M_2}{\sqrt{B}})
\frac{\pm\frac{i}{2}\sqrt{E}-
\frac{1}{2i}\frac{M_3}{\sqrt{C}}}{\frac{1}{2i}(
\frac{M_1}{\sqrt{A}})-i\frac{M_2}{\sqrt{B}})}
&=&0\\
\iff\qquad \sqrt{E}&=&0,\\
\end{eqnarray*}

\noindent which happens only in the uninteresting case ${\bf M}=0$.

Thus each component of $\phi_1$ has $4+8$ poles: four poles
given by the initial data (the poles of $\phi_{0_2}$) and
the eight poles just considered.
But we know that the divisor of $\frac{\phi_{1_2}}{\phi_{1_1}}$ is given by

\begin{equation}\label{ppd}
(\frac{\phi_{1_2}}{\phi_{1_1}})=\epsilon P(\tilde u_1)-P(\tilde u_1)
\end{equation}

\noindent for some $\tilde u_1$ (the same proof as for $\phi_{0_2}$, see above). Whence,
since $all$ poles coincide, eight of
the 12 zeros of $\phi_{1_1}$ coincide with eight zeros
of $\phi_{1_2}$. At those points ${\rm det}\mathcal {V}_0$
must be zero.

Now:

\begin{eqnarray}\label{detV}
{\rm det}{\mathcal V}&=&0\nonumber\\
\iff\quad w_1^2(u_0-u)&=&{\bf V}^{-2}(\frac{-4}{h^2}+
A^{-1}{\bf V}^2-<{\bf V},{\bf I}^{-1}{\bf V}>).
\end{eqnarray}

\noindent $w_1^2(u_0-u)$ is a well defined function on
${\bf T}/\lambda/\epsilon$ and having there only one pole of order
two, it has also two points $u_{1_1}, u_{1_2}$ where (\ref{detV}) is satisfied, thus
in ${\bf T}$ (\ref{detV}) is true at the points $\langle u_{1_1}\rangle\cup
\langle u_{1_2}\rangle$, i.e. there are 16 possible sites for the eight zeros
of $\phi_1$.
Since ${\bf V}$ is a real vector and for small $h$ the
right hand side of (\ref{detV}) converges to $-\infty$, we get (recall the
definition of $w_1$):

\begin{equation}\label{zeros1}
\Im (u_{1_1})=\Im (u_{1_2})=\Im (u_0),
\end{equation}

\noindent and since $w_1$ is odd:

\begin{equation}\label{zeros2}
u_{1_1}+u_{1_2}=2u_0,
\end{equation}

\noindent and for $h\to 0 : u_{1_1},u_{1_2}\to u_0$.
Next, ${\mathcal V}_0\neq 0$ for all $u$, which means that
there is exactly one zero over each element of $\langle u_{1_1}\rangle\cup
\langle u_{1_2}\rangle$.These zeros are invariant w.r.t. $\lambda$ and $\epsilon$:

First note that:

\begin{equation}\label{phiPeriods}
\begin{array}{rclcrcl}
M(\lambda u)&=&\sigma_3M(u)\sigma_3&\Rightarrow &\phi_0(
\lambda p)&=&\sigma_3\phi_0(p),\\
M(\epsilon u)&=&\sigma_1M(u)\sigma_1&\Rightarrow & \phi_0(
\epsilon p)&=&\phi_{0_2}^{-1}(p)\sigma_1\phi_0(p),\\
{\mathcal V}(\lambda u)&=&\sigma_3{\mathcal V}(u)\sigma_3,&&&&\\
{\mathcal V}(\epsilon u)&=&\sigma_1{\mathcal V}(u)\sigma_1.&&&&
\end{array}
\end{equation}

\noindent Thus, if

\begin{equation}\label{pre}
{\mathcal V_0}(pr_{\pi}\,p)\phi_0(p)=0,
\end{equation}

\noindent then

\begin{equation}\label{con}
{\mathcal V_0}(\lambda pr_\pi \,p)\phi_0(\lambda p)={\mathcal V_0}
(\epsilon pr_\pi\, p)\phi_0(\epsilon p)=0.
\end{equation}

\noindent Let $p_{1_1}, p_{1_2}$ be the two zeros with $pr_\pi\, p_{1_i}=
u_{1_i},(i=1,2)$.
We now show that
$p_{1_1}$ and $p_{1_2}$ lie in different sheets: Let $W_0(u):={\mathcal
V_0}(u)-{\bf 1}$, then ${\mathcal V_0}(u_{1_2})={\mathcal V_0}(u_{1_1})
-2W_0(u_{1_1})$, since all $w_k$ are odd.
Next, let ${\mathcal V_0}(u_{1_1})\phi_0(u_{1_1},\mu)=0$. Then:

\begin{eqnarray}\label{different}
{\mathcal V_0}(u_{1_2})\phi_0 (u_{1_2},\mu)&=&({\mathcal V_0}(u_{1_1})-2W_0(u_{1_1}))(
\phi_0(u_{1_1},\mu)+O(h))\nonumber\\
&=&-2W_0(u_{1_1})\phi_0 (u_{1_1},\mu)+O(h)\nonumber\\
&=&2\phi_0 (u_{1_1},\mu)+O(h),\;h\to 0,
\end{eqnarray}

\noindent using the definition of $W_0(u)$. Since $\phi_0 (u_{1_1},\mu)$
doesn't vanish, the second zero of $\phi_1$ must lie on
the other sheet.

Exchanging
the sheets for $p_{1_1}$ and $p_{1_2}$ obviously means changing
the sign of $h$. It is clear that further multiplications
with the matrices ${\mathcal V}_n(u)$ lead to the increase
of the order of the poles over $\langle u_0\rangle$ and to further pairs
of zeros $p_{n_1}$, $p_{n_2}$ common to $\phi_{n_1}$ and $\phi_{n_2}$ near the
points $\langle u_0\rangle$.
The proof that $(\frac{\phi_{n_2}}{\phi_{n_1}})=\epsilon P(\tilde u_n)-P(\tilde u_n)$
for some $\tilde u_n$ is the same as for $\phi_{0_2}$. 

Conversely, given a set of divisors $\{B_n|n\in {\bf N}, n>0\}$ with the above
properties we only have to vary $h\gamma$ for each step to arrive at a sequence
of angular momentum vectors ${\bf M}_n$ which belong to eigenvector functions
with divisors given by the prescribed set of $B_n$ divisors.
\hspace*{\fill}{\bf qed.}
\vspace{.5 cm}

Below we will also need the following properties:

\begin{equation}\label{phiNPeriod1}
\phi_n(\lambda p)=\sigma_3\phi_n(p),
\end{equation}

\noindent since $({\mathcal V}_{n-1}\cdots{\mathcal V}_0\phi_0)(\lambda u)
=(\sigma_3{\mathcal V}_{n-1}\sigma_3\cdots\sigma_3{\mathcal V}_0
\sigma_3\sigma_3\phi_0)(p)=\sigma_3\phi_0(p)$. Similarly:

\begin{equation}\label{phiNPeriod2}
\phi_n(\epsilon p)=\phi^{-1}_{0_2}(p)\sigma_1\phi_n(p).
\end{equation}

\noindent Also recall that the four ``separate'' zeros of $\phi_{n_1}$
and $\phi_{n_2}$ have the symmetry $\chi$:

\begin{equation}\label{zeroSymmetry}
\chi P(\tilde u_n)=P(\tilde u_n).
\end{equation}

\begin{prop}
If $\psi_n$ and $\phi_n$ are two holomorphic vector functions
on the spectral curve which both have divisors as in proposition
(\ref{divisorProp}) and satisfy (\ref{phiNPeriod1}), (\ref{phiNPeriod2}),
and (\ref{zeroSymmetry}), then $\phi_n
=c\psi_n,\; c\in {\bf C}^*$.
\end{prop}

{\bf Proof}: Let $\phi_{n_1},\;\psi_{n_1}$ be the first
coefficients of the vector functions. Then $\frac{\phi_{n_1}}{
\psi_{n_1}}$ is a function on the torus $C/\lambda$
with two poles at $D=p+\chi p$ for some $p$. Two of the three
holomorphic differentials on this compact curve are ${\rm d}u$
and $\frac{{\rm d}u}{\mu}$ (here $u$ is the
pullback of the chart $u$ from ${\bf T}$ to $C$). Its values
in the chart $u$ at the poles are given by:

\begin{equation}\label{table}
\begin{array}{r|cc}
&{\rm d}u & \frac{{\rm d}u}{\mu}\\ \hline
p& 1&\mu^{-1}\\
\chi p & 1 & -\mu^{-1}
\end{array}
\end{equation}

\noindent for some $\mu\neq 0$. The rank of this matrix is two, thus the index of speciality
of $D$ is $1$. From the Riemann-Roch formula it follows
that the space of holomorphic functions $f$ with $(f)+D$
being integral is one. But the constant functions satisfy
this condition, so they are the only ones. Hence $\frac{\phi_{n_1}}{
\psi_{n_1}}=c=const.\;$.

But from (\ref{phiNPeriod2}) we see that also $\frac{\phi_{n_2}}{
\psi_{n_2}}=c=const.$ (the same constant $c$ as for the first
coefficients). Thus $\phi_n=c\psi_n$.\hspace*{\fill} {\bf qed.}

\section{The Prym Theta Function}

In this section the construction of a Prym theta function is described.
This will then be used in the formulas for $\phi_n$.

\begin{figure}
\begin{center}

\begin{picture}(0,0)%
\epsfig{file=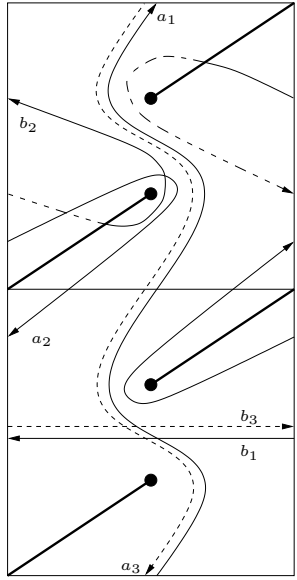}%
\end{picture}%
\setlength{\unitlength}{1973sp}%
\begingroup\makeatletter\ifx\SetFigFont\undefined%
\gdef\SetFigFont#1#2#3#4#5{%
  \reset@font\fontsize{#1}{#2pt}%
  \fontfamily{#3}\fontseries{#4}\fontshape{#5}%
  \selectfont}%
\fi\endgroup%
\begin{picture}(3666,7266)(2368,-7594)
\put(4276,-586){\makebox(0,0)[lb]{\smash{\SetFigFont{6}{7.2}{\rmdefault}{\mddefault}{\updefault}$a_1$}}}
\put(3826,-7486){\makebox(0,0)[lb]{\smash{\SetFigFont{6}{7.2}{\rmdefault}{\mddefault}{\updefault}$a_3$}}}
\put(2701,-4636){\makebox(0,0)[lb]{\smash{\SetFigFont{6}{7.2}{\rmdefault}{\mddefault}{\updefault}$a_2$}}}
\put(2551,-1936){\makebox(0,0)[lb]{\smash{\SetFigFont{6}{7.2}{\rmdefault}{\mddefault}{\updefault}$b_2$}}}
\put(5326,-5611){\makebox(0,0)[lb]{\smash{\SetFigFont{6}{7.2}{\rmdefault}{\mddefault}{\updefault}$b_3$}}}
\put(5326,-6061){\makebox(0,0)[lb]{\smash{\SetFigFont{6}{7.2}{\rmdefault}{\mddefault}{\updefault}$b_1$}}}
\end{picture}

\caption{appropriate homology basis for $\frac{E}{M^2}>B^{-1}$} \label{homologyBasisEGreater}
\end{center}
\end{figure}

Consider the Riemann surface $C/\lambda$. We will first investigate the
case $\frac{E}{M^2}>B^{-1}$. The homology basis is chosen as
in figure (\ref{homologyBasisEGreater}).

\noindent That is 

\begin{equation}\label{PrymHomology}
\begin{array}{rclcrcl}
\pi a_1&=&-a_3&,&\pi b_1&=&-b_3,\\
\pi a_2&=&-a_2&,&\pi b_2&=&-b_2,
\end{array}
\end{equation}

\noindent The corresponding normalized cohomology basis will be denoted by
$v_k,\; (k=1,2,3)$:

\begin{equation}\label{normalization}
\int_{a_k}v_l=2\pi i \delta_{kl}\; .
\end{equation}

\noindent A Prym differential for the projection $\pi$ is given by

\begin{equation}\label{omega}
\omega:=v_1+v_3\; .
\end{equation}

\noindent We then have

\begin{equation}\label{omegaOdd}
\pi^*\omega=\pi^*(v_1+v_3)=-v_3-v_1=-\omega
\end{equation}

\noindent (For a thorough discussion of Prym differentials and Prym varieties
see [4]. In our particular case the differential $\omega$ can be given
explicitly by $\omega=\frac{2\pi i}{c}\frac{{\rm d}u}{\mu}$ where
$c:=\int_{a_1}\frac{{\rm d}u}{\mu}$, but this fact won't be used.)

\begin{lemma}\label{iLemma}
We have

\begin{eqnarray}
&&\int_{a_1}\omega=\int_{a_3}\omega=2\pi i, \int_{a_2}\omega=0,\label{i}\\
&&\int_{b_1}\omega=\int_{b_3}\omega=:\tau,\label{ii}\\
&&\int_{b_2}\omega=2\pi i.\label{iii}
\end{eqnarray}
\end{lemma}

{\bf Proof}: The assertions (\ref{i}) are obvious from the definition of
$\omega$. Further:

\begin{equation}\label{omegaB}
\int_{b_3}\omega=\int_{-\pi b_1}\omega=-\int_{b_1}\pi^*\omega=\int_{b_1}\omega .
\end{equation}

Finally, to prove (\ref{iii}), first note that

\begin{equation}\label{ba}
b_2+\epsilon b_2=a_1+a_3,
\end{equation}

\begin{figure}
\begin{center}

\begin{picture}(0,0)%
\epsfig{file=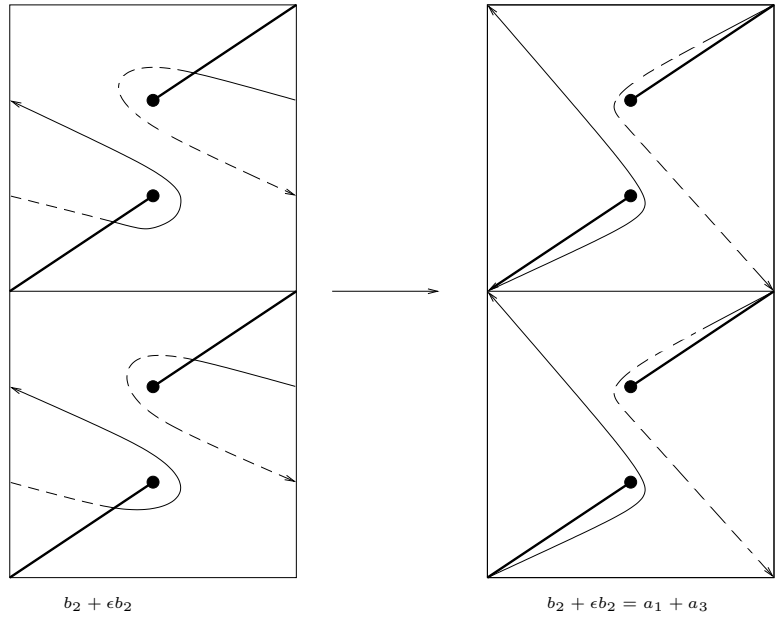}%
\end{picture}%
\setlength{\unitlength}{1973sp}%
\begingroup\makeatletter\ifx\SetFigFont\undefined%
\gdef\SetFigFont#1#2#3#4#5{%
  \reset@font\fontsize{#1}{#2pt}%
  \fontfamily{#3}\fontseries{#4}\fontshape{#5}%
  \selectfont}%
\fi\endgroup%
\begin{picture}(9666,7666)(1168,-7994)
\put(1876,-7936){\makebox(0,0)[lb]{\smash{\SetFigFont{6}{7.2}{\rmdefault}{\mddefault}{\updefault}$b_2+\epsilon b_2$}}}
\put(7951,-7936){\makebox(0,0)[lb]{\smash{\SetFigFont{6}{7.2}{\rmdefault}{\mddefault}{\updefault}$b_2+\epsilon b_2=a_1+a_3$}}}
\end{picture}

\caption{in the homology group: $b_2+\epsilon b_2=a_1+a_3$} \label{deform1}
\end{center}
\end{figure}

\noindent (see figure (\ref{deform1})).

\noindent Moreover

\begin{equation}\label{epsilonOmega}
\epsilon^*\omega=\omega.
\end{equation}

\noindent This is true, since the $a$-periods of $\epsilon\,\omega$ and $\omega$
coincide:

\begin{eqnarray}
\epsilon\, a_1&=&a_1,\label{eA1}\\
\epsilon\, a_3&=&a_3,\label{eA3}
\end{eqnarray}

\noindent so the $a_1$- and $a_3$-periods coincide. Because of

\begin{equation}\label{epsilonA}
a_2+\epsilon\, a_2=0,
\end{equation}

\begin{figure}
\begin{center}

\begin{picture}(0,0)%
\epsfig{file=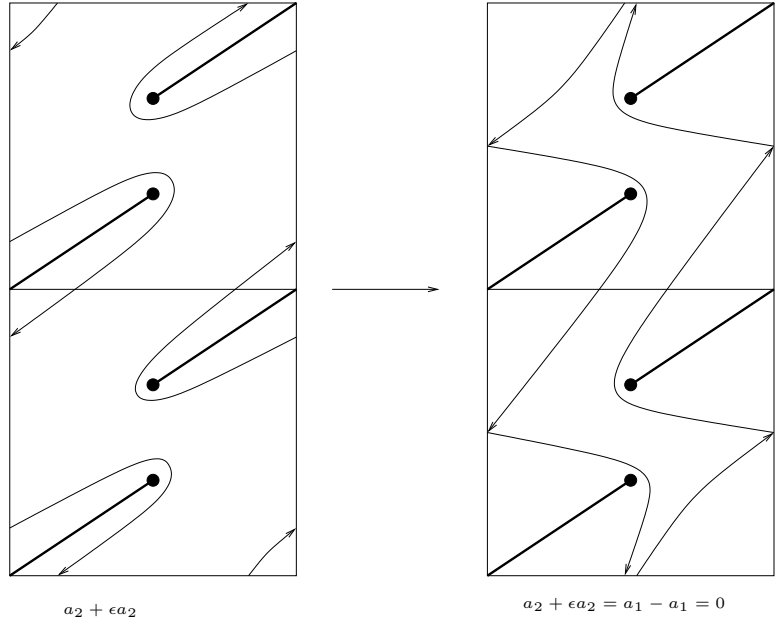}%
\end{picture}%
\setlength{\unitlength}{1973sp}%
\begingroup\makeatletter\ifx\SetFigFont\undefined%
\gdef\SetFigFont#1#2#3#4#5{%
  \reset@font\fontsize{#1}{#2pt}%
  \fontfamily{#3}\fontseries{#4}\fontshape{#5}%
  \selectfont}%
\fi\endgroup%
\begin{picture}(9666,7741)(1168,-8069)
\put(1876,-8011){\makebox(0,0)[lb]{\smash{\SetFigFont{6}{7.2}{\rmdefault}{\mddefault}{\updefault}$a_2+\epsilon a_2$}}}
\put(7651,-7936){\makebox(0,0)[lb]{\smash{\SetFigFont{6}{7.2}{\rmdefault}{\mddefault}{\updefault}$a_2+\epsilon a_2=a_1-a_1=0$}}}
\end{picture}

\caption{In the homology group $a_2+\epsilon a_2=0$} \label{deform2}
\end{center}
\end{figure}

\noindent (see figure (\ref{deform2})),

\noindent we also have:

\begin{equation}\label{null}
\int_{a_2}\epsilon^*\omega=-\int_{\epsilon\,a_2}\epsilon^*\omega=-\int_{a_2}\omega=0,
\end{equation}

\noindent so indeed $\epsilon^*\omega=\omega$.
Finally, using (\ref{ba}) and (\ref{epsilonOmega}) it follows

\begin{equation}\label{finally}
\int_{b_2}\omega=\frac{1}{2}(\int_{b_2}\omega+\int_{\epsilon\,b_2}
\epsilon^*\omega)=\frac{1}{2}\int_{a_1+a_3}\omega=2\pi i.
\end{equation}
\hspace*{\fill}{\bf qed.}
\vspace{.5 cm}

Next, the torus given by the two generators $(2\pi i,\,\tau)$ will be denoted by
$J$. Let $p_0$ be an arbitrary point on $C/\lambda$, then

\begin{equation}\label{abel}
A(p):=\int_{p_0}^p\omega
\end{equation}

\noindent is a well defined map from $C/\lambda$ to $J$ because of lemma
(\ref{iLemma}). The theta function on $J$ is defined by:

\begin{equation}\label{theta}
\theta(z,\tau):=\sum_{m\in {\bf Z}}\exp(\frac{1}{2}m^2\tau+mz).
\end{equation}

\noindent The periods of $\theta(z,\tau)$ are given by ($e\in J$, arbitrary):

\begin{eqnarray}
\theta(z+e+2\pi i)&=&\theta(z+e)\label{thetaPeriod1}\\
\theta(z+e+\tau)&=&\exp(-\frac{\tau}{2}-z-e)\theta(z+e),\label{thetaPeriod2}
\end{eqnarray}

\noindent from which the periods of the pullback $\theta(A(p)+e)$ of
$\theta(z+e)$ to $C/\lambda$ follow.

\begin{prop}\label{thetaProp}
The function $\theta(A(p)+e)$ has two zeros in $C/\lambda$ and if $\breve p$
is one of them, then $\chi\,\breve p$ is the other.
\end{prop}

{\bf Proof}:
The proof that $\theta(A(p)+e)$ has two zeros in $C/\lambda$ follows the
standard procedure: just cut $C$ along the homology basis and then integrate
$\frac{{\rm d}\theta(z+e)}{\theta(z+e)}$ along the image of that cut in $J$.

To prove the second part of the proposition, first note, that
$\chi^*\omega=\omega$: this follows from:

\begin{equation}\label{chiA}
\begin{array}{rcl}
\chi a_1&=&a_3,\\
\chi a_2&=&-a_2.
\end{array}
\end{equation}

\noindent So let's choose $p_0$ to be one of the points over $u=0$.
Let $\breve p$ be a zero of $\theta(A(p)+e)$ and $\zeta$ a path from
$p_0$ to $\breve p$, i.e. $\chi\, \zeta$ is a path from $p_0$ to
$\chi \breve p$. Then:

\begin{equation}\label{thetaZeros}
A(\breve p)=\int_{\zeta}\omega=\int_{\chi\zeta}\chi^*\omega=\int_{\chi\zeta}
\omega=A(\chi\,\breve p),
\end{equation}

\noindent which proves the second part.\hspace*{\fill}{\bf qed.}
\vspace{.5 cm}

\begin{figure}[h]
\begin{center}

\begin{picture}(0,0)%
\epsfig{file=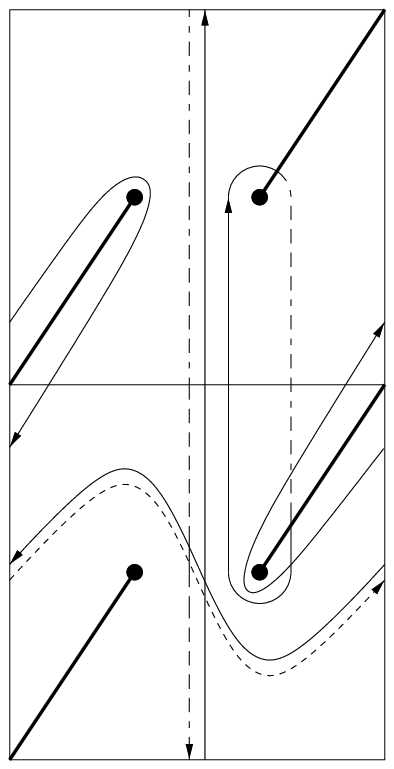}%
\end{picture}%
\setlength{\unitlength}{1973sp}%
\begingroup\makeatletter\ifx\SetFigFont\undefined%
\gdef\SetFigFont#1#2#3#4#5{%
  \reset@font\fontsize{#1}{#2pt}%
  \fontfamily{#3}\fontseries{#4}\fontshape{#5}%
  \selectfont}%
\fi\endgroup%
\begin{picture}(4050,7770)(3226,-7861)
\put(5401,-286){\makebox(0,0)[lb]{\smash{\SetFigFont{6}{7.2}{\rmdefault}{\mddefault}{\updefault}$a_1$}}}
\put(3226,-5686){\makebox(0,0)[lb]{\smash{\SetFigFont{6}{7.2}{\rmdefault}{\mddefault}{\updefault}$b_1$}}}
\put(5026,-7861){\makebox(0,0)[lb]{\smash{\SetFigFont{6}{7.2}{\rmdefault}{\mddefault}{\updefault}$a_3$}}}
\put(7276,-6136){\makebox(0,0)[lb]{\smash{\SetFigFont{6}{7.2}{\rmdefault}{\mddefault}{\updefault}$b_3$}}}
\put(3751,-4561){\makebox(0,0)[lb]{\smash{\SetFigFont{6}{7.2}{\rmdefault}{\mddefault}{\updefault}$a_2$}}}
\put(5701,-1711){\makebox(0,0)[lb]{\smash{\SetFigFont{6}{7.2}{\rmdefault}{\mddefault}{\updefault}$b_2$}}}
\end{picture}

\caption{appropriate homology basis for $\frac{E}{M^2}<B^{-1}$} \label{homologyBasisESmaller}
\end{center}
\end{figure}

For the case $\frac{E}{M^2}<B^{-1}$ we choose the homology basis given
in figure (\ref{homologyBasisESmaller}). With this choice, everything
works as in the other case above.

\section{The Baker-Akhiezer Function}
Now we formulate the main theorem. Let's recall the notations:

Let ${\bf M_0}\in {\bf R}^3$ be an initial value for the discrete Euler top,
$C$ the spectral curve defined by ${\bf M_0}$ and let $P(\breve u)$
be the pole divisor of $\phi_{0_2}$ ($\phi_0$ is the eigenvector function
of $M_0$ with $\phi_{0_1}=1$). $\omega$ and $\theta$ will still denote the
Prym differential and theta function described above, and 

\begin{equation}\label{smallE}
e:=-\left(\int^{(\breve u,\alpha(\breve u))}\omega\right) + \pi i+\frac{\tau}{2}.
\end{equation}

\noindent Further, to each $n\in {\bf N}$ we associate the divisors
$B_n$, $E_n$ as in (\ref{b}) and (\ref{e}). On the curve $C/\lambda/\epsilon$
let ${\rm d}\Omega_n$ be the third kind abelian differential with zero
$a$-periods (the $a$-cycles on $C/\lambda/\epsilon$ (let's denote them
by $\tilde a$) are defined by:

\begin{eqnarray}
2\tilde a_1&=&pr_\epsilon\, a_1\label{aTilde1}\\
2\tilde a_2&=&pr_\epsilon \,a_3\label{aTilde2})
\end{eqnarray}

\noindent and only single poles with residues given by the divisor $E_n$

\begin{equation}\label{residues}
res_p{\rm d}\Omega=E_n(p),\; p\in C/\lambda/\epsilon.
\end{equation}

\noindent Finally,

\begin{equation}\label{V}
V_n=\int_{b_1}{\rm d}\Omega_n.
\end{equation}

\noindent In what follows, pullbacks of differentials are usually denoted with
the same symbol as the original differentials and it will be clear from the context
which one is meant.

\begin{theorem}\label{TT}
The functions:

\begin{eqnarray}
\phi_{n_1}(p)&:=&\frac{\theta(\int^p\omega+\sum_{k=1}^n V_k+e)}
{\theta(\int^p\omega+e)}\exp(\int^p\sum_{k=1}^n {\rm d}\Omega_k)\label{phiN1}\\
\phi_{n_2}(p)&:=&\frac{\theta(\int^p\omega+\sum_{k=1}^n V_k+\pi i+e)}
{\theta(\int^p\omega+e)}\exp(\int^p\sum_{k=1}^n {\rm d}\Omega_k)\label{phiN2}
\end{eqnarray}

are well defined on $C$ and are the coefficient functions of an eigenvector
function of $M_n$.
\end{theorem}

{\bf Proof}: We first prove that $\phi_{n_1}$ and $\phi_{n_2}$ are functions
on $C$. Let's consider $C/\lambda$. The $a$-cycles cause no problems. For the
$b_2$-cycle, note that

\begin{equation}\label{bOmegaNull}
\int_{b_2}{\rm d}\Omega_k=\frac{1}{2}(\int_{b_2}{\rm d}\Omega_k+
\int_{\epsilon b_2}\epsilon^* {\rm d}\Omega_k)=\frac{1}{2}\int_{a_1+a_3}
{\rm d}\Omega_k=0.
\end{equation}

Next we show that $\int_{b_1}{\rm d}\Omega_k=\int_{b_3}{\rm d}\Omega_k$.
Since ${\rm d}\Omega_k$ is a normalized abelian differntial of the third
kind, we have the Riemann-Roch relations:

\begin{eqnarray}
\int_{b_1}{\rm d}\Omega_k&=&\int_{E_n+\epsilon E_n}v_1,\label{rbr1}\\
\int_{b_3}{\rm d}\Omega_k&=&\int_{E_n+\epsilon E_n}v_3,\label{rbr2}
\end{eqnarray}

\noindent Now $v_1-v_3=\frac{2\pi i}{4iK'}{\rm d}u$, since they obviously
have the same $a$-periods. But

\begin{equation}\label{du}
\int_{E_n+\epsilon E_n}{\rm d}u=0
\end{equation}

\noindent and thus the $b_1$- and $b_3$-periods of  ${\rm d}\Omega_k$
are the same.
Hence $\phi_{n_1}$ does not change w.r.t. any cycle in $C/\lambda$ while
$\phi_{n_2}$ changes sign when going around $b_1$ and $b_3$ in $C/\lambda$,
i.e. on $C$ it is single valued too.

Now we proceed to show that $\phi_n=(\phi_{n_1},\phi_{n_2})^t$ is indeed
an eigenvector function of $M_n$.
Recall that we have to show the following facts:
\begin{description}
\item[(i)] both $\phi_{n_1}$ and $\phi_{n_2}$ have four poles given by $P(\breve u)$
(determined by the initial value),
\item[(ii)] both $\phi_{n_1}$ and $\phi_{n_2}$ have poles and zeros given by the
divisor $\sum_{k=1}^n B_k$,
\item[(iii)] the remaining four zeros in the divisors of $\phi_{n_1}$ and $\phi_{n_2}$
are not the same but in both cases they are symmetric w.r.t. the involution
$\chi$,
\item[(iv)] $\phi_n(u+2K,\mu)=\sigma_3 \phi_n(u,\mu)$,
\item[(v)] $\phi_n(u+2iK',\mu)=\phi_{0_2}^{-1}(u,\mu)\sigma_1 \phi_n(u,\mu)$.
\end{description}

The fact (i) follows from the appropriate choice of the point $e$ in the torus
$J$. $\phi_{n_1}$ and $\phi_{n_2}$ thus both have a pole at $(\breve u,\alpha(
\breve u))$. From proposition \ref{thetaProp} and the behaviour of $\theta$ when
shifted by $\tau$, we find that $\theta(A(p)+e)$ has zeros at $P(\breve u)$.

Point (ii) is provided by the exponential terms.

Next, we know already that the second coefficient of an eigenvector function 
of a matrix $M$ of the form $\phi=(1,\phi_2)^t$ has a divisor symmetric w.r.t.
the involution $\chi$. Now the pole and zero divisors of  $\phi_n=(1,\phi_{n_2})^t$
are given by the four zeros of $\phi_{n_1}$ and $\phi_{n_2}$ respectively, which
proves (iii).

(iv) follows immediately from the periodic behaviour of $\theta$.

Finally, since
$\omega$ as well as ${\rm d}\Omega_k$ are pullbacks via the quotient map
$C/\lambda\to C/\lambda/\epsilon$, and $a_1$- and $a_3$-periods for $\omega$
are $2\pi i$ and for ${\rm d}\Omega_k$ zero, we get:

\begin{eqnarray}
\lefteqn{\phi_{n_1}(u+2iK',\mu)=}\nonumber\\
&=&\frac{\theta(\int^{(u,\mu)}\omega +\sum V_k+\pi i+e)}
{\theta(\int^{(u,\mu)}\omega+\pi i+e)}\exp(\int^{(u,\mu)}\sum{\rm d}\Omega_k),\label{1}\\
&&\nonumber\\
\lefteqn{\phi_{n_2}(u+2iK',\mu)=}\nonumber\\
&=&\frac{\theta(\int^{(u,\mu)}\omega +\sum V_k+2\pi i+e)}
{\theta(\int^{(u,\mu)}\omega+\pi i+e)}\exp(\int^{(u,\mu)}\sum{\rm d}\Omega_k).\label{2}
\end{eqnarray}

\noindent Hence:

\begin{eqnarray}
\phi_n(u+2iK',\mu)&=&\frac{\theta(\int^{(u,\mu)}\omega +e)}
{\theta(\int^{(u,\mu)}\omega+\pi i+e)}\;\sigma_1\phi_n(u,\mu)\nonumber\\
&=&\phi_{0_2}^{-1}(u,\mu)\sigma_1 \phi_n(u,\mu),
\end{eqnarray}

\noindent which proves (v).\hspace*{\fill}{\bf qed.}
\vspace{.5 cm}

\section{Solutions For The Angular Momentum}

To get ${\bf M}$ from the Baker-Akhiezer functions, we consider (\ref{auxil})
at the point $u=0$. Here the asymptotics of $M_n$ and $\mu$ are given by

\begin{eqnarray}
M_n&=&(\frac{\rho}{2i}\sum M_k\sigma_k)u^{-1}+O(1),\; u\to 0,\label{MAsymp}\\
\mu&=&\pm (\frac{\rho}{2i}|M|)u^{-1}+O(1),\; u\to 0,\label{muAsymp}
\end{eqnarray}

\noindent Using as base $p_0$ for the integrals a point with $pr_\pi\, p_0=0$,
we see that the two eigenvectors at $u=0$ can, after multiplication with
appropriate factors, be written as

\begin{equation}\label{PhiAt0}
\Phi_n(p_0)=\left(
\begin{array}{cc}
1&1\\
\frac{\theta(\pi i+\sum V_k+e)}{\theta(\sum V_k+e)}&
\frac{\theta(d+\pi i+\sum V_k+e)}{\theta(d+\sum V_k+e)}
\end{array}\right)
=:
\left(
\begin{array}{cc}
1&1\\A&B
\end{array}\right),
\end{equation}

\noindent where $d:=\int_{p_0}^{\pi p_0}\omega$. Using this,
(\ref{auxil}) leads to

\begin{equation}\label{scratch}
\left(
\begin{array}{cc}
M_3+(M_1-iM_2)A & M_3+(M_1-iM_2)B\\
M_1+iM_2-M_3A & M_1+iM_2-M_3B
\end{array}
\right)
=
\left(
\begin{array}{cc}
M & -M\\
AM & -BM
\end{array}\right).
\end{equation}

\noindent This finally yields

\begin{equation}\label{finalMs}
\begin{array}{rcl}
M_1&=&M\frac{\theta(\sum V_n+e)\theta(d+\sum V_n+e)-
\theta(\pi i+\sum V_n+e)\theta(d+\pi i+\sum V_n+e)}{
\theta(\pi i+\sum V_n+e)\theta(d+\sum V_n+e)-
\theta(\sum V_n+e)\theta(d+\pi i+\sum V_n+e)}\label{finalM1}\\
&&\\
M_2&=&iM\frac{\theta(\pi i+\sum V_n+e)\theta(d+\pi i+\sum V_n+e)+
\theta(\sum V_n+e)\theta(d+\sum V_n+e)}{
\theta(\pi i+\sum V_n+e)\theta(d+\sum V_n+e)-
\theta(\sum V_n+e)\theta(d+\pi i+\sum V_n+e)}\label{finalM2}\\
&&\\
M_3&=&M\frac{\theta(\pi i+\sum V_n+e)\theta(d+\sum V_n+e)+
\theta(\sum V_n+e)\theta(d+\pi i+\sum V_n+e)}{
\theta(\sum V_n+e)\theta(d+\pi i+\sum V_n+e)-
\theta(\pi i+\sum V_n+e)\theta(d+\sum V_n+e)}.\label{finalM3}
\end{array}
\end{equation}

\section{Acknowledgements}
I want to thank A.Bobenko, R.Seiler, and J.Suris for lots of helpful
dis-cussions.

\section{Bibliography}

[1] A.I.Bobenko, B.Lorbeer, Y.Suris. Integrable discretizatons of
the Euler top, Preprint No. 312, SFB 288 "Differentialgeometrie und
Quantenphysik"

\noindent [2] E.D.Belokolos, A.I.Bobenko, V.Z. Enol'ski, A.R. Its, V.B. Matveev.
Al-gebro-geometric approach to nonlinear integrable equations, Springer,
Berlin (1994)

\noindent [3] B. Dubrovin, I. Krichever, S.Novikov. Integrable Systems I, In:
Dynamical systems IV. Encyclopaedia of mathematical sciences (vol.4),
V.I. Arnol'd, S.P. Novikov (Eds.)

\noindent [4] J.Fay. Theta functions on Riemann surfaces, Lecture Notes in Mathematics
352, Springer, Berlin (1973)

\noindent [5] M.Ablowitz, J. Ladik. A nonlinear difference scheme
and inverse scattering. Stud. Appl. Math. 55 (1976) 213-229; On the
solution of a class of nonlinear partial difference equations.
Stud. Appl. Math. 57 (1977) 1-12.

\noindent [6] F.W. Nijhoff, H.Capel, and G.Wiersma. Integrable lattice
systems in two and three dimensions. Lect. Notes Phys. 239 (1984)
263-302; Linearizing integral transform for the multicomponent lattice
KP. Physica A 138 (1986) 76-99.

\noindent [7] A.P. Veselov. Integrable systems with discrete time
and difference operators. Funct.Anal. Appl. 22 (1988) 1-13.

\noindent [8] Yu.B. Suris. Generalized Toda Chains in discrete time.
Algebra i Anal. 2 (1990) 141-157; Discrete-time generalized Toda
lattices: complete integrability and relation with relativistic
Toda lattices. Phys. Lett. A 145 (1990) 113-119.

\noindent [9] Discrete Integrable Geometry and Physics, Eds. A.Bobenko
and R.Seiler, Oxford University Press, 1998.

\noindent [10] J.Moser, A.P. Veselov. Discrete Versions of some
classical integrable systems and factorization of matrix polynomials.
Commun. Math. Phys. 139 (1991) 217-243.

\noindent [11] P.Deift, L.-Ch.Li, and C.Tomei. Loop groups, discrete
versions of some classical integrable systems, and rank 2 etensions.
Mem. Amer. Math. Soc. 479 (1992).

\noindent [12] A.I. Bobenko. Real algebro-geometric solutions of the
Landau-Lifshits equation in Prym theta functions. Functional Anal.
Appl. 19, 5-17 (1985).

\end{document}